\documentstyle[aps,prl,epsfig,ifthen,multicol]{revtex}

\begin{document}

\title{Ferromagnetism in magnetically doped III-V semiconductors}
\author{V.~I.~Litvinov$^{1}$
\cite{email}, V.~K.~Dugaev$^{2,3}$}
\address{$^1$WaveBand Corporation, 375 Van Ness ave., Suite 1105,
Torrance, California 90501, USA\\
$^2$Max-Planck-Institut f\"ur Mikrostrukturphysik,
Weinberg 2, D-06120 Halle, Germany\\
$^3$Institute for Materials Science Problems, 
Ukrainian Academy of Sciences,\\ 
Vilde 5, 58001 Chernovtsy, Ukraine}
\date{Received \today }
\maketitle

\begin{abstract}
The origin of ferromagnetism in semimagnetic III-V materials is 
discussed. The indirect exchange interaction caused by virtual 
electron excitations from magnetic impurity level in the bandgap to 
the valence band can explain ferromagnetism in GaAs(Mn) no matter 
samples are degenerated or not. Formation of ferromagnetic clusters 
and percolation picture of phase transition describes well all 
available experimental data and allows to predict the Mn-composition 
dependence of transition temperature in wurtzite (Ga,In,Al)N epitaxial 
layers.
\end{abstract}

\begin{multicols}{2}

Electronic and optoelectronic semiconductor devices, controlled 
by a weak magnetic field, and electric-field controlled ferromagnetism
in semiconductors\cite{ohno} promise new functionality of memory, 
detector and light emitting sources.  Possible device implementations 
of spin-electronics are high-electron-mobility transistors, Si/Si 
and GaAs/Si spin-valve transistors, spin light-emitting diodes, quantum 
computers, and integration of nonvolatile storage and logic. Efficiency 
of spin injection depends on a quality of interfaces, and all-semiconductor 
structures should benefit the performance of spin-electronic devices. 
Recent achievement in material research resulted in ferromagnetic 
semiconductor material lattice-matched to III-V semiconductors: highest 
ever ferromagnetic critical temperature in semiconductors ($T_c=110$~K) has 
been observed in metallic samples of Ga$_{1-x}$Mn$_x$As                      
($x=0.053$)\cite{matsukura,beschoten}. 

The origin of ferromagnetism in III-V materials is not well understood. 
The explanation of ferromagnetism by holes-mediated 
Ruderman-Kittel-Kasuya-Yosida (RKKY) interaction (Ref.\cite{matsukura}) 
gives theoretical critical temperature $T_c$ pretty close to that 
observed experimentally. 
A detailed theory of the ferromagnetism based on RKKY interaction, has been 
found in recent articles\cite{dietl,macdonald}. 

Despite the efforts, some difficulties arise in free carrier-based explanation 
of ferromagnetism in GaAs(Mn). First, nonzero $T_c$ has been observed 
in the low-carrier concentration samples, where holes are not degenerated. 
Second, holes density in GaAs increases with Mn content, thus what 
we see as carrier density dependence of $T_c$ might be the dependence on 
a localized spin concentration. Third, the estimation, made in 
Ref.\cite{matsukura}, employs the mean-field approximation, which is 
valid only when the interaction radius $L$ is much larger than the average 
interspin distance $\overline{R}$. 
In the samples under consideration in Ref.\cite{matsukura} this is 
not a case since $L$ was estimated as short due to crystal imperfections,   
$L\simeq (5-6.5)$~\AA \, whereas Mn content $x=0.053$ corresponds to
$\overline{R}=6$~\AA .
 
Free carriers contribute to the ferromagnetism in cubic III-V samples with 
degenerated holes, but until now it is not clear if this contribution 
is responsible for high $T_c$ or not. In wide-bandgap semiconductors the 
degenerated free carriers can hardly be obtained and RKKY-based explanation 
is also under the question. Recent prediction of room temperature 
ferromagnetism in wide-bandgap GaN(Mn)\cite{dietl} is based on an 
assumption that crystal is degenerated. However, in real $p$-type Mn- or 
Fe- doped GaN samples holes are not degenerated. 

Our point is that the RKKY mechanism is not an ultimate reason of 
ferromagnetism and high ferromagnetic transition temperature can occur 
even in non-degenerate semiconductors. In this paper, we discuss the 
alternative mechanism of ferromagnetism in III-V materials doped with 
magnetic atoms. A localized spin in a crystal excites band electrons due 
to $s$-$p$ or $p$-$d$ exchange interaction and generally gives rise to three types 
of indirect exchange interaction between impurity spins caused by virtual 
excitations of band electrons. If the Fermi level lies inside an energy gap 
there is a threshold for electron excitations, and the indirect exchange 
drops exponentially with a distance between impurities. The energy gap 
determines the length of the exponential decay (Bloembergen-Rowland 
mechanism\cite{bloembergen}). 
In indirect-gap semiconductors electron excitations involve the momentum 
transfer ${\bf K}$. This modulates the exponential decay 
by oscillations with the period $\sim K^{-1}$ (Ref.\cite{abrikosov80}).  
If the Fermi level is inside the band, the excitations 
of electrons near Fermi momentum $\hbar k_F$ result in the range function that 
oscillates with period $\sim k_F^{-1}$ and the amplitude falling as a power of a 
distance. This is the long-ranged RKKY interaction\cite{rkky}.

In this paper we would like to attract attention to excitations of 
Mn acceptors, namely, to virtual acceptor level-valence band transitions 
rather than electron-hole pair excitations around Fermi level in RKKY 
theory. This mechanism works no matter the sample is degenerated or not 
and it could also be a reason of ferromagnetism in GaAs(Mn) as well as 
in wide bandgap materials like GaN(Mn). In non-degenerated sample, where 
Fermi level lies in the bandgap, this mechanism is the only possible one to 
mediate an interaction between localized spins. 

The Bloembergen-Rowland-type indirect interaction between two 
magnetic ions, separated by the distance $r$, is given as\cite{litvinov}:
$$
J(r)=-\frac{J_{pd}^2\, m^2\, \Delta }{4\pi ^3\, \hbar ^4\, n^2\, r^2}
K_2\left( 2r/r_0\right) \; ,\; \; \; \; 
r_0=\hbar \, (m\Delta )^{-1/2}\; .
\eqno (1)
$$
Here $J_{pd}$ is the exchange interaction between free carrier and localized 
spin, $n$ is the concentration of host atoms in the sublattice of substitution 
(cation sublattice in GaAs), $m$ is the reduced electron mass in the bands 
under consideration, $\Delta $ is the smallest energy gap for electron excitations, 
and $K_2(y)$ is the McDonald function. The indirect exchange interaction, Eq.~(1), 
tends to known results in two limiting cases: 

a) small distances (narrow-gap semiconductors)\cite{liu}, $r\ll r_0\; $, 
$J(r)\sim 1/r^4$ , 

b) large distances\cite{abrikosov80}, $r\gg r_0\; $,\\ 
$J(r)\sim r^{-5/2}\exp (-2r/r_0)$.

In a pure semiconductor, the smallest energy gap $\Delta $ is the bandgap. 
We will use the expression Eq.~(1) in the limit when one of the band is a
narrow impurity band formed by Mn or Fe in highly doped material. In 
the non-degenerated $p$-GaAs(Mn) the smallest gap is the activation energy 
of a hole at Mn atom $\Delta \simeq 113$~meV\cite{schraier}, the reduced 
mass is the valence band density of state mass $m\simeq 0.53\, m_0$. 
The characteristic length for GaAs(Mn) is $r_0\simeq 11$~\AA . 

\begin{figure}
\hskip0.7cm \noindent
\epsfig{file=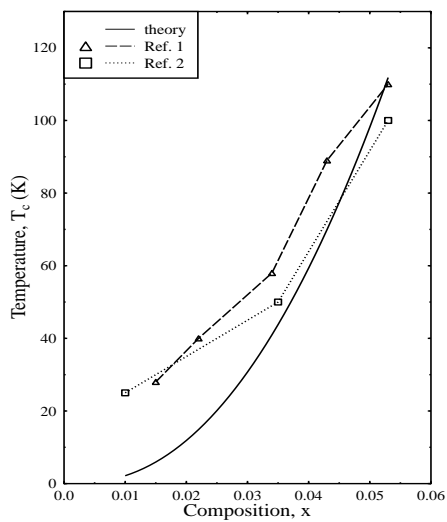,height=9cm,width=7cm}
\caption{Mn concentration dependence of ferromagnetic transition temperature 
in GaAs(Mn).}
\end{figure}

As the interaction is short-ranged, the mean field approximation is not 
helpful for the calculation of the ferromagnetic critical temperature $T_c$.  
Instead, we have to use the percolation approach\cite{korenblit}. 
At certain temperature, 
the spins, coupled by a strong exchange interaction, form a ferromagnetic 
cluster. The cluster size $R_{cl}$ is determined by the equation 
$S(S+1)J(R_{cl})=k_BT$. When 
temperature decreases, the clusters grow in size due to coalescence, and, 
when a percolation threshold is reached, the 'infinite' cluster penetrates 
the whole crystal. The percolation threshold determines the transition 
temperature and it is reached when cluster size becomes
$R_{cl}=R_{perc}\simeq \overline{R}\, (B_c)^{1/3}$, where $B_c$ 
is a geometrical factor. In the three-dimensional random-site model, 
this factor is given as $B_c=2.4$ (Ref.\cite{holcomb}). 
The ferromagnetic critical temperature follows: 
$$
k_B\, T_c=S\, (S+1)\, J(R_{perc})\; .
\eqno (2)
$$

The composition dependence of $T_c$ comes from the average interspin 
distance, which in GaAs(Mn) has the form $\overline{R}=(3a^3/16\pi x)^{1/3}$. 
We used $p$-$d$ exchange interaction $J_{pd}=2.5$~eV\cite{szczytko}.  
Results of $T_c$ calculation along with available experimental data 
are shown in Fig.~1.

Increase in Mn concentration beyond $x=0.055$ most likely causes compensation, 
the Fermi level moves toward the conduction band and then appears to lie 
between the Mn acceptor level and a conduction band. At this point, the 
smallest gap is the energy difference between the acceptor level and the 
edge of the conduction band. This difference is large, that makes the 
exchange interaction small and $T_c$ low (see Eq.~(1)). 

\begin{figure}
\hskip0.7cm \noindent
\epsfig{file=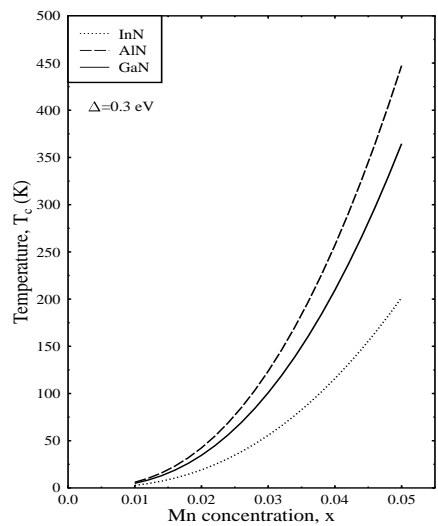,height=9cm,width=7cm}
\caption{Ferromagnetic transition temperature in wurtzite group 
III-Nitrides doped with Mn.}
\end{figure}

The prediction of the impurity band formation in GaN(Mn,Fe) was reported 
in Ref.~\cite{schilfgaarde}. 
In $p$-type group-III Nitrides the acceptor levels lie in the bandgap, not 
closer than 200-250~meV to the valence band. The average interspin distance 
in wurtzite GaN(Mn) is $\overline{R}=(3^{4/3}a^2\, c/16\pi \, x)^{1/3}$,
where $a$ and $c$ are the lattice constants.  
$P$-$d$ exchange interaction is unknown and can be estimated the same way as 
in Ref.~\cite{dietl}: the $J_{pd}$ value in GaAs is multiplied by the 
ratio of cation densities in wurtzite GaN and cubic GaAs. 
This ratio is found as 1.95 (GaN), 1.45 (InN), and 2.163 (AlN). 
The calculations are done for $m=0.8\,m_0$, and  the results are shown 
in Fig.~2.

\begin{figure}
\hskip0.7cm \noindent
\epsfig{file=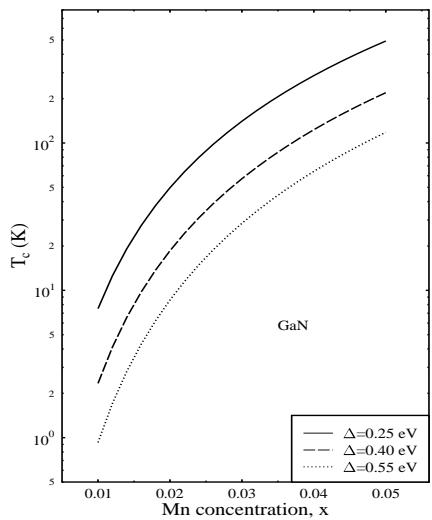,height=9cm,width=7cm}
\caption{Concentration dependence of the transition temperature for different 
values of the energy gap.}
\end{figure}

The transition temperature, shown in Fig.~2, is sensitive to the value 
of the minimum energy gap $\Delta $. At present, there is no much information 
about actual position of Mn or Fe levels in the bandgap of GaN-based 
materials. According to local density approximation (LDA) results for 
cubic GaN (Fe,Mn)\cite{fong}, Mn gives rise to two peaks of density of states with 
one portion merging with top of the valence band. Because the accuracy 
of LDA results is of the order of 1~eV, experimental studies are also needed to 
specify the position of Mn and Fe levels in the bandgap. Fig.~3 illustrates 
the $\Delta $-dependence of $T_c$ in wurtzite GaN.

Since ferromagnetic ordering appears in a crystal, a spin polarization 
of free carriers can reveal itself in many different experiments, 
no matter whether ferromagnetism is caused by free electrons or not. 
So the mechanism, discussed here, is not in contradiction with 
anomalous Hall effect\cite{matsukura} and magnetic dichroism in optical 
absorption\cite{beschoten}, observed in GaAs(Mn).

In conclusion, we have discussed the specific contribution to indirect 
exchange interaction in III-V-cubic and wurtzite materials that is 
caused by virtual impurity level-valence band electron excitations. 
As the indirect exchange interaction is short-ranged, the mean-field theory 
is not valid and we use more adequate percolation picture of phase 
transition. The contribution we calculated can explain ferromagnetism 
in GaAs(Mn) and possible ferromagnetism in wide bandgap materials. 
As far as absolute value of $T_c$ is concerned, it depends on the values of $p$-$d$ 
exchange interaction $J_{pd}$ and the energy gap between impurity and band 
electron levels $\Delta $, which are uncertain to some extent, especially 
for GaN material system. Further progress in understanding of impurity 
ferromagnetism in III-V materials will depend mostly on growth and magnetic 
characterization of epitaxial GaN-InN-AlN(Mn,Fe), and further experimental 
studies of GaAs(Mn).

We acknowledge J.~Barna\'s for reading the manuscript and useful comments. 
V.L. thanks H.~Morkoc for fruitful discussions. 
V.D. is grateful to U.~Krey for stimulating discussions and P.~Bruno
for the encouragement.

\end{multicols}

\begin{references}

\bibitem[*]{email}
E-mail: vlitvinov@earthlink.net

\bibitem{ohno}
H.~Ohno, D.~Chiba, F.~Matsukura, T.~Omlya, E.~Abe, T.~Dietl, Y.~Ohno,
and K.~Ohtani, Nature {\bf 408}, 944 (2000).

\bibitem{matsukura}
F. Matsukura, H. Ohno, A. Shen, and Y. Sugawara, 
Phys. Rev. B {\bf 57}, R2037 (1998); 
H.~Ohno and F.~Matsukura, Sol. State Commun. {\bf 117}, 179 (2001).

\bibitem{beschoten}
B.~Beschoten, P.~A.~Crowell, I.~Malajovich, D.~D.~Awschalom, 
F.~Matsukura, A.~Shen, and H.~Ohno,  
Phys. Rev. Lett. {\bf 83}, 3073 (1999).

\bibitem{dietl}
T.~Dietl, A.~Haury, and Y.~M.~d'Aubign\'e, Phys. Rev. B {\bf 55}, 
R3347 (1997);
T.~Dietl, H.~Ohno, F.~Matsukara, J.~Cibert, and D.~Ferrand, 
Science {\bf 287}, 1019 (2000);
T.~Dietl and H.~Ohno, cond-mat/0002450;
T.~Dietl, H.~Ohno, and F.~Matsukura, cond-mat/0007190;

\bibitem{macdonald}
T.~Jungwirth, W.~A.~Atkinson, B.~H.~Lee, and A.~H.~MacDonald, Phys. Rev. B 
{\bf 61}, 15606 (2000); 
J.~K\"onig, H.-H.~Lin, and A.~H.~MacDonald, 
Phys. Rev. Lett. {\bf 84}, 5628 (2000);
J.~Schliemann, J.~K\"onig, H.-H.~Lin, and A.~H.~MacDonald, cond-mat/0010036;
J.~K\"onig, H.-H.~Lin, and A.~H.~MacDonald, cond-mat/0010471;
J.~Schliemann, J.~K\"onig, and A.~H.~MacDonald, cond-mat/0012233;

\bibitem{bloembergen}
N.~Bloembergen and T.~Rowland, Phys. Rev. {\bf 97}, 1679 (1955).

\bibitem{abrikosov80}
A.~A.~Abrikosov, J. Low Temp. Phys. {\bf 39}, 217 (1980).

\bibitem{rkky}
M.~A.~Ruderman and C.~Kittel, Phys. Rev. {\bf 96}, 99 (1954); 
T.~Kasuya, Prog. Theor. Phys. {\bf 16}, 45 (1956); 
K.~Yosida, Phys. Rev. {\bf 106}, 893 (1957).

\bibitem{litvinov}
V.~I.~Litvinov, Fiz. Tekh. Poluprov. {\bf 19}, 555 (1985)
[Sov. Phys. Semicond. {\bf 19}, 345 (1985)].

\bibitem{liu}
L.~Liu and G.~Bastard, Phys. Rev. B {\bf 52}, 487 (1982).

\bibitem{schraier}
W.~Schraier and M.~Schmidt, Phys. Rev. B {\bf 10}, 2501 (1974).

\bibitem{korenblit}
I.~Ya.~Korenblit and E.~F.~Shender, Uspekhi Fiz. Nauk {\bf 126}, 233 
(1978) [Sov. Phys. Uspekhi {\bf 21}, 832 (1978)];   
B.~I.~Shklovskii and A.~L.~Efros, 
{\em Electronic properties of doped semiconductors} (Springer, 
New York, 1984). 

\bibitem{holcomb}
D.~F.~Holcomb and J.~J.~Rehr, Phys. Rev. {\bf 183}, 773 (1966);  
D.~F.~Holcomb, M.~Iwasawa, and F.~D.~K.~Roberts, Biometrica {\bf 59}, 
207 (1972).

\bibitem{szczytko}
J.~Szczytko, W.~Mac, A.~Stachow, A.~Twardowski, P.~Becla, and J.~Tworzydlo, 
Solid State Commun. {\bf 99}, 927 (1996).

\bibitem{schilfgaarde}
M.~Van~Schilfgaarde and O.~Mryasov, APS Meeting, abstract V26.006, 
Minneapolis MN, March 20-24, 2000.

\bibitem{fong}
C.~Y.~Fong, V.~A.~Gubanov, and C.~Boekema,  
J. Electron. Mater. {\bf 29},1067 (2000).

\end{references}
\end{document}